\documentclass[sigconf]{acmart}

\AtBeginDocument{%
  \providecommand\BibTeX{{%
    \normalfont B\kern-0.5em{\scshape i\kern-0.25em b}\kern-0.8em\TeX}}}

\usepackage{multirow,array}
\usepackage{makecell}
\usepackage{comment}
\usepackage{enumitem} % itemize no indent
\usepackage{tcolorbox}
\usepackage{booktabs}
\usepackage{xspace}
\usepackage{url}
\usepackage{subcaption}
\usepackage{listings}
\usepackage{xcolor}
\usepackage{tikz}
\usepackage{amsmath}
\usetikzlibrary{shapes, arrows.meta, positioning}

\newcommand{\tool}{{\textsf{PerfSense}}\xspace}

\newcommand{\dev}{{\textsf{DevAgent}}\xspace}
\newcommand{\perf}{{\textsf{PerfAgent}}\xspace}

\newcommand{\DiagConfig}{{\textsf{DiagConfig}}\xspace}
\newcommand{\ChatGPT}{{\textsf{ChatGPT}}\xspace}

\newcommand{\phead}[1]{\vspace{1mm} \noindent {\bf #1}}

\usepackage{tcolorbox}

\tcbset{
    colback=gray!10, % 设置背景色为白色
    colframe=gray, % 设置边框颜色为灰色
    coltext=black, % 设置文本颜色为黑色
    boxsep=5pt, % 内边距
    arc=0mm, % 边角的弧度
    top=0mm, bottom=0mm, % 上下边距
    left=1mm, right=1mm, % 左右边距
    boxrule=0.5mm, % 边框的粗细
    toptitle=1mm, bottomtitle=1mm, % 标题上下的空白
    titlerule=0mm, % 标题下方的线条粗细
}
\definecolor{deepgreen}{rgb}{0.0, 0.4, 0.0}

\definecolor{codegreen}{rgb}{0,0.6,0}
\definecolor{codegray}{rgb}{0.5,0.5,0.5}
\definecolor{codepurple}{rgb}{0.58,0,0.82}
\definecolor{backcolour}{rgb}{0.95,0.95,0.92}

\lstdefinestyle{mystyle}{
  backgroundcolor=\color{backcolour}, commentstyle=\color{codegreen},
  keywordstyle=\color{magenta},
  numberstyle=\tiny\color{codegray},
  stringstyle=\color{codepurple},
  basicstyle=\ttfamily\footnotesize,
  breakatwhitespace=false,         
  breaklines=true,                 
  captionpos=b,                    
  keepspaces=true,                 
  numbers=left,                    
  numbersep=5pt,                  
  showspaces=false,                
  showstringspaces=false,
  showtabs=false,                  
  tabsize=2
}
\newtcolorbox[auto counter,number within=chapter]{definition}[1][]{
  enhanced,
  breakable,
  fonttitle=\scshape,
  title={Definition \thetcbcounter},
  #1
}

\begin{document}

\title{Identifying Performance-Sensitive Configurations in Software Systems through Code Analysis with LLM Agents}
%Leveraging LLM-Based Agents to Identify Performance-Sensitive Configurations in Software Systems

% \affiliation{%


\begin{abstract}

Configuration settings are essential for tailoring software behavior to meet specific performance requirements. However, incorrect configurations are widespread, and identifying those that impact system performance is challenging due to the vast number and complexity of possible settings. In this work, we present \tool, a lightweight framework that leverages Large Language Models (LLMs) to efficiently identify performance-sensitive configurations with minimal overhead. \tool employs LLM agents to simulate interactions between developers and performance engineers using advanced prompting techniques such as prompt chaining and retrieval-augmented generation (RAG). Our evaluation of seven open-source Java systems demonstrates that \tool achieves an average accuracy of 64.77\% in classifying performance-sensitive configurations, outperforming both our LLM baseline (50.36\%) and the previous state-of-the-art method (61.75\%). Notably, our prompt chaining technique improves recall by 10\% to 30\% while maintaining similar precision levels. Additionally, a manual analysis of 362 misclassifications reveals common issues, including LLMs' misunderstandings of requirements (26.8\%). In summary, \tool significantly reduces manual effort in classifying performance-sensitive configurations and offers valuable insights for future LLM-based code analysis research.

\end{abstract}




\author{Zehao Wang, Dong Jae Kim, Tse-Husn (Peter) Chen \\Software PErformance, Analysis and Reliability (SPEAR) Lab\\ Concordia University, Montreal, Canada \\ w\_zeha@encs.concordia.ca, k\_dongja@encs.concordia.ca, peterc@encs.concordia.ca}
\renewcommand{\shortauthors}{Zehao Wang, Dong Jae Kim, and Tse-Husn (Peter) Chen}

\maketitle
\section{Introduction} \label{intro}
Modern software systems feature numerous configuration options, enabling customization for diverse workloads and hardware platforms~\cite{singh2016optimizing,bao2018autoconfig}. While these configurations provide flexibility, some configurations, known as performance-sensitive configurations, can impact system performance when their values change. Developers need to identify and understand the impact of such configurations to ensure they are set correctly, maintaining system performance and behavior. However, due to the large volume of configurations, pinpointing performance-sensitive configurations is time-consuming~\cite{jin2012understanding,han2016empirical} and incorrect settings are a common source of system misbehavior and performance degradation~\cite{ganapathi2004pcs, LinkedInComm}. 
Hence, automated approaches to quickly find performance-sensitive configurations that require special attention or further investigation are important to alleviate developers burden~\cite{Yonkovit2015,tian2015latency}.

Performance experts have various tools at their disposal to assess performance-sensitive configurations. Alongside performance profiling tools~\cite{Jprofiler,visualvm,bornholt2018finding}, they can identify inefficient code patterns~\cite{chen2014detecting, liu2014characterizing,nistor2015caramel}, and utilize data-flow and dynamic analysis to find performance-sensitive configurations~\cite{li2020statically,lillack2014tracking}. However, as highlighted by ~\citet{velez2022debugging}, the adoption of these tools faces usability challenges for performance experts when analyzing the performance impact of configurations. These challenges arise from (1) a lack of comprehensive understanding of the codebase and its intricate interactions across multiple components, (2) difficulties in identifying the code affected by performance-sensitive configurations, and (3) the intricate cause-and-effect relationship between performance-sensitive configurations and the corresponding source code. Consequently, performance engineers may face challenges in accurately identifying the performance sensitivity of configurations. 
Effective collaboration between developers and performance engineers is crucial for overcoming these challenges and effectively identifying performance-sensitive configurations. Developers possess in-depth knowledge about the codebase and its functionality, while performance engineers specialize in analyzing performance-related issues. Leveraging their complementary expertise enables more thorough code analysis and more accurate classification of performance-sensitive configurations. 

The rise of Large Language Models (LLMs) is revolutionizing programming and software engineering. Trained on vast code datasets, LLMs understand code deeply and excel in various code-related tasks. With tools like ChatGPT~\cite{chatGPT} and LLaMA~\cite{LLAMA}, researchers showcase LLMs' potential in tasks like generating commit messages~\cite{zhang2024automatic}, resolving merge conflicts~\cite{shen2023git}, creating tests~\cite{xie2023chatunitest,yuan2023no,schafer2023empirical}, renaming methods~\cite{alomar2024refactor}, and aiding in log analytics~\cite{ma2024knowlog,ma2024llmparser}. Given the complexity of collaboration during software engineering tasks, using LLM agents stands out as a promising direction to replicate human workflows. Specifically, multi-agent systems have achieved significant progress in solving complex tasks by assigning agents to specific roles and emulating collaborative activities in software engineering practice~\cite{hong2023metagpt,dong2023self,qian2023communicative}. For example, \citet{dong2023self} developed a self-collaboration framework, assigning LLM agents to work as distinct experts for sub-tasks in software development. \citet{qian2023communicative} proposed an end-to-end framework for software development through self-communication among the agents. 

Inspired by multi-agent, we introduce \tool, a lightweight framework designed to effectively classify performance-sensitive configurations using Large Language Models (LLMs) as multi-agent systems.
\tool leverages the collaborative capabilities of LLMs to mimic the interactions between developers and performance engineers, enabling a thorough analysis of the performance sensitivity of configurations. 
\tool employs two primary agents: \dev and \perf. \dev focuses on retrieving relevant source code and documentation related to the configurations and conducting performance-aware code reviews. \perf, on the other hand, utilizes the insights from \dev to classify configurations based on their performance sensitivity. This collaboration is facilitated through advanced prompting techniques such as prompt chaining and retrieval-augmented generation (RAG), which enhance the agents' understanding and analytical capabilities.

To address the challenge of navigating a large codebase with limited LLM context size, \tool iteratively breaks down complex tasks into manageable subtasks. Specifically, \perf iteratively communicates with \dev to gather and analyze relevant source code associated with the configurations under scrutiny. Through a series of prompt chains, \perf refines its understanding by requesting specific details, clarifications, and performance-related insights from \dev. This iterative communication ensures that \perf accumulates a comprehensive knowledge base without exceeding the context size limitations, enabling accurate classification of performance-sensitive configurations.

Our evaluation of seven open-source systems demonstrates that \tool achieves 64.77\% accuracy in classifying performance-sensitive configurations, outperforming state-of-the-art technique~\cite{10.1145/3611643.3616300} and our LLM baseline with an average accuracy of 61.75\% and 50.36\%, respectively. 
Compared to prior technique~\cite{10.1145/3611643.3616300} that requires tens or hundreds of hours to collect performance data manually, \tool is lightweight and requires minimal human effort.

In summary, we make the following contributions:

\begin{itemize}
    \item Our evaluation of seven open-source systems demonstrates that \tool achieves an average accuracy of 64.77\%, surpassing the state-of-the-art approaches with an average accuracy of 61.75\%.
    \item We proposed a new LLM-based code analysis technique that employs two primary agents, \dev and \perf, to navigate large codebases with limited LLM context sizes through advanced prompting techniques such as prompt chaining and retrieval-augmented generation (RAG).
    %\item \textbf{Balanced Precision and Recall}: SensitiveTeeth achieves a better balance of precision (61.47\%) and recall (83.95\%) compared to baselines, ensuring a comprehensive approach to identifying performance-sensitive configurations.
    \item We analyzed the effect of different prompting components that we implemented in \tool. We found that our prompt chaining technique significantly improves the recall (10\% to 30\% improvement) while maintaining a similar level of precision. 
    \item We conducted a manual study of the 362 misclassified configurations, identifying key reasons for misclassification, including LLM's misunderstanding of requirements (26.8\%) and incorrect interpretation of performance impact (10.0\%). 
    \item We provided a discussion on the implications of our findings and highlight future direction on LLM-based code analysis. 
\end{itemize}

%In conclusion, \tool advances the state-of-the-art in performance-sensitive configuration identification, providing a practical and efficient solution for developers and performance engineers. This leads to better software performance and reduced costs associated with misconfiguration.
In conclusion, %\tool represents a meaningful step forward in the field of performance-sensitive configuration identification. B
by leveraging multi-agent collaboration and advanced prompting techniques, \tool provides an efficient technique for classifying performance-sensitive configuration, one of the most important first steps in understanding system performance. \tool also presents a novel code navigation approach that may inspire future LLM-based research on code analysis. %solution for developers and performance engineers, contributing to improved software performance and helping to mitigate costs associated with performance misconfigurations.

\phead{Paper Organization.} Section~\ref{sec:background} provides the background of the problem and technique. Section~\ref{sec:related} discusses related work. Section~\ref{sec:method} presents the details of \tool. 
Section~\ref{sec:evaluate} shows the evaluation results. 
Section~\ref{sec:disscussion} discusses the findings.
Section~\ref{sec:threats} discusses the threats to validity. 
Section~\ref{sec:conclusion} concludes the paper. 

%\djk{add what we did and result}

%In summary, we make the following contributions:
%\begin{enumerate}
%    \item XX
    
%\end{enumerate}

\section{Background}
\label{sec:background}
In this section, we first discuss the definition and importance of performance-sensitive configuration. Then, we provide background on large language models (LLM) agents and retrieval-augmented generation (RAG).

\subsection{Performance-Sensitive Configurations} 
Software systems often contain various configuration parameters to provide flexibility in deployment and execution~\cite{singh2016optimizing,bao2018autoconfig}. Some configurations, known as performance-sensitive configurations, affect performance when their values change. For example, an application's name is generally not performance-sensitive, whereas memory allocation settings can significantly impact performance~\cite{yin2011empirical, cacheoptimizer}. Identifying these configurations is crucial, as their usage directly impacts system efficiency and stability. However, developers may not always be aware of the performance implications of configuration changes, leading to common misconfigurations, impacting overall system performance~\cite{yin2011empirical, 10.1145/2517349.2522727, cacheoptimizer, 10.1145/2961111.2962602, 10.1145/3510003.3510043}.

Determining which configurations are performance-sensitive is challenging, given the high number of configurations and complex interactions among various system components~\cite{zhang2015performance}, the absence of transparent documentation or feedback concerning the performance implications of each setting~\cite{yin2011empirical}, and the complexity and time-intensive nature of performance testing~\cite{Yonkovit2015}. 
Performance engineers need to conduct load tests to evaluate the performance sensitivity and impacts of various configurations. {\em These tests involve altering the values of configuration parameters and assessing their impacts on system performance~\cite{zhang2015performance,singh2016optimizing,wang2021would,vitui2021mlasp}}. 
Therefore, an important step that can reduce the testing cost is to only conduct such tests on performance-sensitive configurations.

%\subsection{Collaborative Efforts in Identifying Performance-Sensitive Configurations} 
While developers implement code functionality with the best coding standards in mind, they may not always adhere to best-performance engineering practices. In collaborative efforts, developers and performance engineers work together to identify performance-sensitive configurations. Performance engineers leverage domain-specific knowledge to design and implement performance tests that uncover configuration sensitivities. However, performance engineers need the assistance of developers who have an in-depth understanding of the codebase to navigate across multiple source code components. Hence, to narrow down performance-sensitive configurations that impact overall system performance, there must be synergy in sharing knowledge between developers and performance engineers. %Identifying performance-sensitive configurations is crucial to ensure optimal system performance, reduce resource utilization, improve user response time, and potentially save companies billions of dollars~\cite{Yonkovit2015}.

\subsection{LLM-based Multi-agent Framework} 

Large language models (LLMs) are pre-trained using vast datasets comprising a wide range of texts, such as documentation and source code. %Recently, LLMs have been shown to generate responses that are similar to human-level quality~\cite{shen2023git,liu2024make,zhang2024automatic}. In particular, with tools like ChatGPT~\cite{chatGPT} or LLaMA~\cite{LLAMA}, researchers have demonstrated the potential of LLMs in generating commit messages~\cite{zhang2024automatic}, resolving merge conflicts~\cite{shen2023git}, generating tests~\cite{xie2023chatunitest,yuan2023no,schafer2023empirical}, method renaming~\cite{alomar2024refactor}, and facilitating log analytics~\cite{ma2024knowlog,ma2024llmparser}. 
The core of LLM agents consists of large language models (LLMs) designed to understand questions and generate human-like responses. These agents refine their responses based on feedback~\cite{madaan2024self}, use memory mechanisms to learn from historical experiences~\cite{li2024personal}, retrieve informative knowledge to improve prompting and generate better responses~\cite{zhao2023retrieving}, and collaborate with other LLM agents to solve complex tasks in a multi-agent process~\cite{guo2024large}. By using prompting, agents can assume specific roles (e.g., developer or tester) and provide domain-specific responses~\cite{deshpande2023anthropomorphization}. In particular, a multi-agent system has been shown to improve the capabilities of individual LLM agents by enabling collaboration among agents, each with specialized abilities~\cite{hong2023metagpt,chan2023chateval}. Multiple LLM agents can share domain expertise and make collective decisions. Effective communication patterns are crucial for optimizing the overall performance of a multi-agent framework, allowing them to tackle complex projects using a divide-and-conquer approach~\cite{chen2023agentverse}. Finally, with modern frameworks like LangChain~\cite{langlgraph}, one key characteristic of LLM agents is their ability to interact with external tools to perform tasks similarly to humans. For example, an LLM agent acting as a test engineer can generate test cases, use test automation tools to collect code coverage, and answer further queries based on the gathered information. 

In this paper, we propose \tool, which leverages LLM agents to emulate the collaboration between developers and performance engineers. \tool analyzes the source code and classifies whether a configuration is performance-sensitive. \tool is zero-shot and unsupervised. It requires minimal input from developers and achieves better results than the state-of-the-art technique on classifying performance-sensitive configuration~\cite{10.1145/3611643.3616300}. 

% \peter{This part is too short. Check Feng's paper}\peter{you need to mention 1) what is an agent (role-playing, interaction, etc) 2) what does an agent have (e.g., memory) 3) what does an agent need (external data, using tools or RAG)} \peter{you need to mention}LLM agents extend the capabilities of traditional LLMs by incorporating decision-making and interaction abilities within specific domains or applications. These agents are designed to perform tasks that require not only understanding and generating text but also integrating external data or APIs. \peter{don't mention langraph here}LangGraph~\cite{langlgraph} is designed to support the creation and management of LLM-based agents within a multi-agent system, where each agent can be specialized to handle specific tasks or domains of knowledge.

% Finally, with modern frameworks like LangChain~\cite{langlgraph}, LLM agents can interact with external tools to perform tasks similarly to humans. For example, an LLM agent acting as a test engineer can generate test cases, use test automation tools to collect code coverage, and answer further queries based on the gathered information. 

% The tools within LangGraph refer to the functionalities and methods that enable these agents to perform tasks such as APIs, functions, and databases. Users can define their tools, allowing agents to select the appropriate tool to complete specific tasks. For example, a user wants to 

\section{Related Work} \label{sec:related}
%This section provides an overview of the publications relevant to our problem under study. 
In this section, we discuss existing research and literature on three topics: 1) Performance Analysis of Configuration; 2) Using LLMs to Analyze Configuration; and 3) Multi-Agent-Based Code Analysis. %similar topics, exploring different approaches and methodologies used by researchers to address similar challenges.

\subsection{Performance Analysis of Configuration}
Some previous research aims to analyze the performance of configuration to help developers understand the performance issue during the software configuration tuning. ConfigCrusher~\cite{10.1007/s10515-020-00273-8} relies on static taint analysis to reveal the relationship between an option and the affected code regions, dynamically analyze the influence of configuration options on the regions' performance, and build the performance-influence model through white-box performance analysis. DiagConfig~\cite{10.1145/3611643.3616300} leverages static taint analysis to identify the dependencies between performance-related operations and options. Through manual performance experiments and labeling on training systems, they build a random forest model to classify the performance-sensitive configurations. Different from the above work, in our work, we employ the LLM agent alongside static code analysis, specifically the call graph analysis, to study the performance of configurations. Given LLMs' promising performance in understanding code, call graph analysis for LLMs can provide more information and incur lower overhead compared to taint analysis.
More importantly, our approach is zero-shot and reduces minimal human effort, which can help developers efficiently identical potential performance-sensitive configurations for further analysis. 

\subsection{Using LLMs to Analyze Configuration}
Recently, large language models have shown promising performance on various software engineer tasks, such as code generation and summarization. Much research leveraged LLMs for tasks related to software configuration. ~\citet{lian2024configuration} proposed an LLM-based framework, Ciri, using few-shot learning and prompt engineering to validate the correctness of configuration files from the file level and parameter level. From the evaluation of real-world misconfigurations, comprising 64 configurations, and synthesized misconfigurations involving 1,582 parameters, Ciri achieves F1 scores of 0.79 and 0.65 at the file level and parameter level, respectively.~\citet{liu2024llmcompdroid} introduced the LLM-CompDroid framework, which employs LLMs alongside the bug resolution tool to address configuration compatibility issues in Android applications. Their framework surpasses the state-of-the-art (SOTA) methods by at least 9.8\% and 10.4\% in the Correct and Correct@k metrics, respectively, respectively.~\citet{shi2024logconfiglocalizer} came up with the framework, LogConfigLocalizer, which leverages Large Language Models and logs to localize root-cause configuration properties, achieving a high average accuracy of 99.91\%. Different from these works, our work explores the potential of LLMs to analyze the performance sensitivity of configurations, which can assist developers in reducing performance testing costs.

\subsection{Multi-Agent Based Code Analysis}

Agent-based code analysis emphasizes the importance of defining roles and facilitating communication among multiple LLM agents. Some approaches incorporate external tools as agents. For example, \citet{huang2023agentcoder} introduced a test executor agent that employs a Python interpreter to provide test logs for LLMs. Similarly, \citet{zhong2024ldb} presented a debugger agent that uses a static analysis tool to construct control flow graphs, aiding LLMs in locating bugs. Other studies~\cite{hong2023metagpt,qian2023communicative,dong2023self, lin2024llm} assigned LLMs to emulate diverse human roles, such as analysts, engineers, testers, project managers, and chief technology officers (CTOs). These approaches use software process models (e.g., Waterfall) for inter-role communication, varying the prompts and roles to enhance code generation. %Many of these benchmark utilize common benchmarks like \textit{Humaneval}~\cite{chen2021evaluating}, \textit{Humaneval-ET}~\cite{dong2023codescore}, \textit{MBPP}~\cite{austin2021program}, \textit{MBPP-ET}~\cite{liu2023your} to measure efficiency of their techniques. 
Our technique leverages similar multi-agent systems to classify performance-sensitive configurations. By integrating prompt chaining and retrieval-augmented generation (RAG), \tool enhances the collaborative capabilities of LLM agents, leading to a lightweight technique that addresses the challenge of limited LLM context size when analyzing a large codebase.   %more accurate and efficient code analysis.

\begin{figure*}
  \centering 
    %\vspace{-3mm}
  \scalebox{0.85}{
      \includegraphics[width=\linewidth]{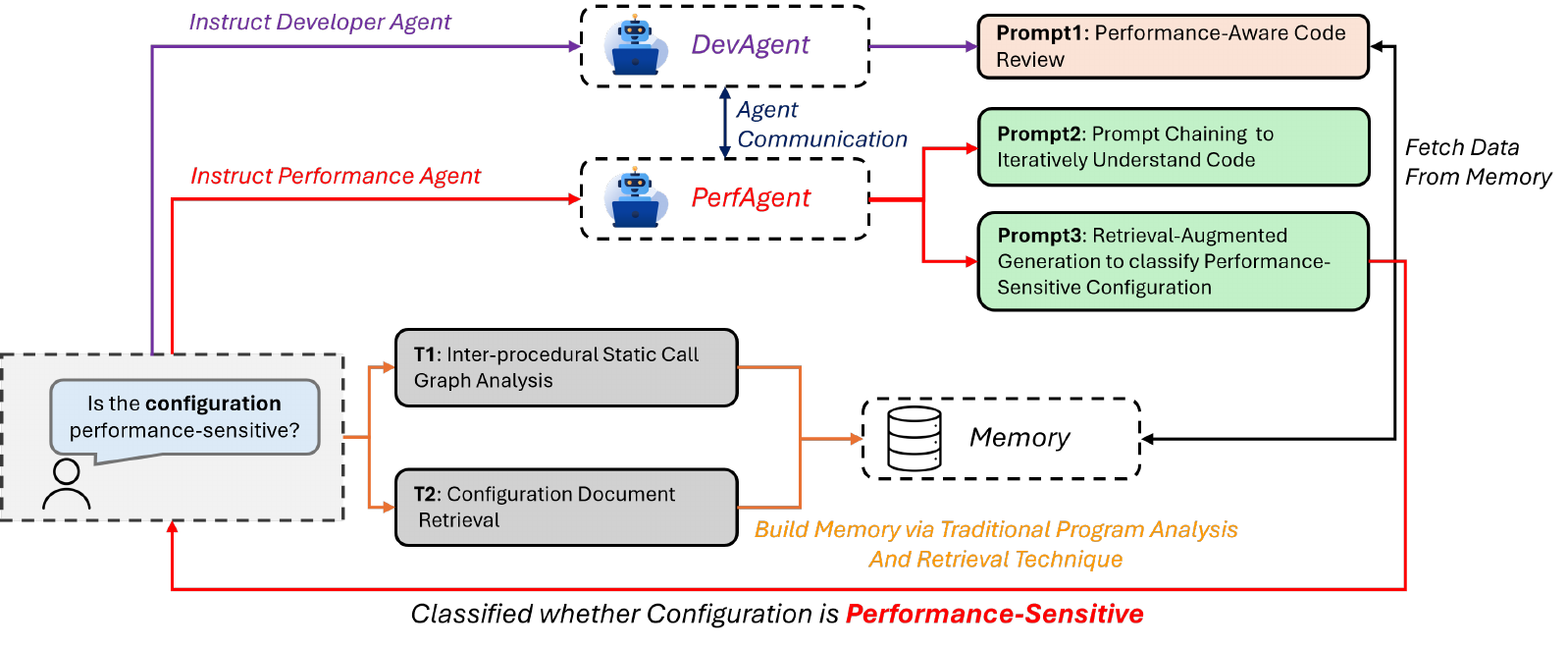}
      }
        %\vspace{-3mm}
  \caption{Overview of \tool}
  \label{fig:overview}
    \vspace{-1.5mm}
\end{figure*}

\section{Design of \tool} \label{sec:method}
In this section, we introduce \tool, a lightweight framework designed for identifying performance-sensitive configurations. We begin by discussing various LLM agents and their communication and conclude with a detailed running example.

%\subsection{An Overview of the Multi-Agent Components} 
Figure~\ref{fig:overview} illustrates the overview of \tool. To analyze the performance sensitivity of configurations, \tool comprises two different agents: the developer (\textbf{\dev}) and the performance expert (\textbf{\perf}). At a high level, given a potential performance-sensitive configuration, \perf utilizes iterative self-refinement and retrieval-augmented prompting techniques in a zero-shot setting, with the assistance of \dev, to iteratively build a knowledge base of the codebase and classify whether the configuration is performance-sensitive. In the following section, we elaborate on the roles of \perf and \dev, and their communication pattern for determining performance-sensitive configurations.

\subsection{Agent Roles and Definition} 
\label{agents role}

\begin{figure}
  \centering
  %\vspace{-3mm}
  \includegraphics[width=\linewidth]{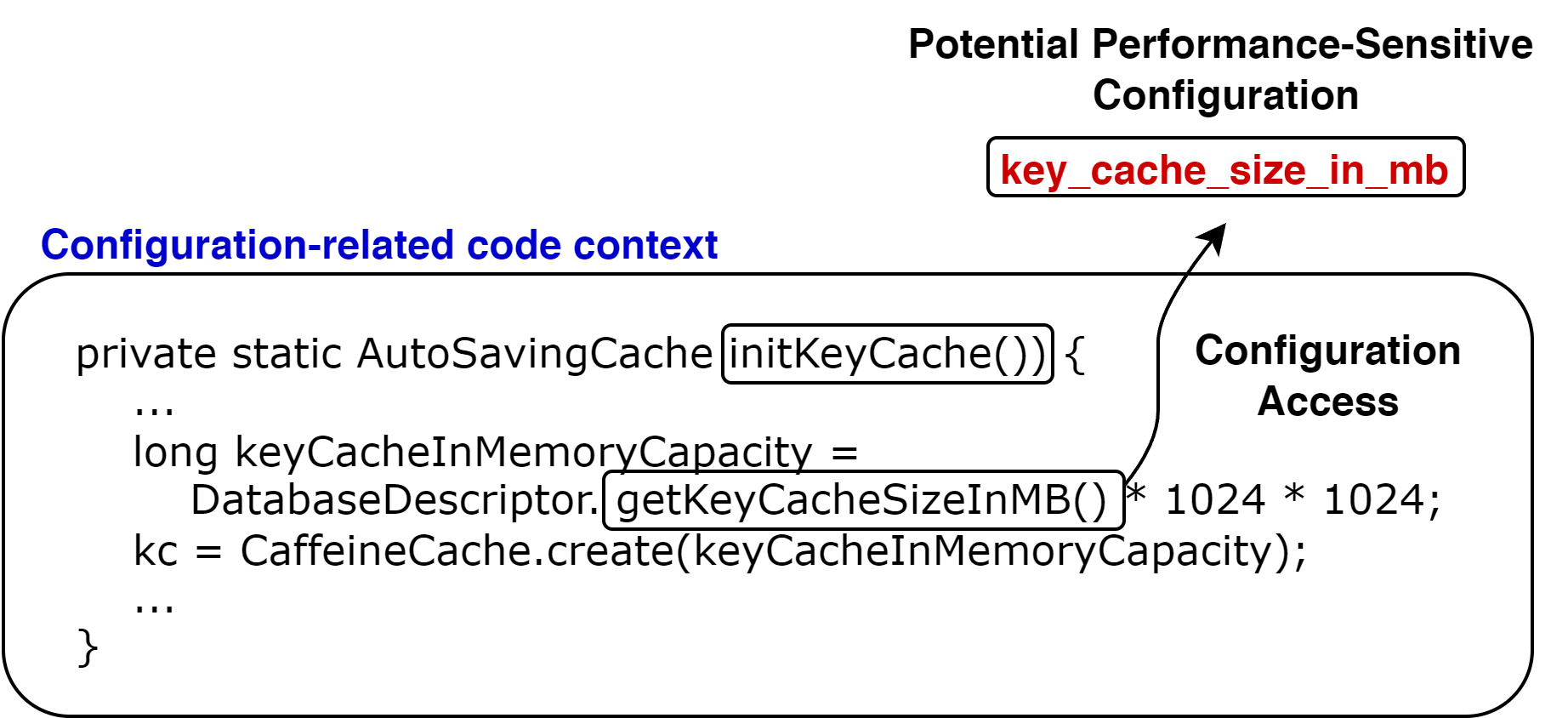}
  \vspace{-5mm}
  \caption{An example of performance-sensitive configuration.}
  \label{fig:background}
  \vspace{-3mm}
\end{figure}

\subsubsection{\textbf{\underline{Developer Agent}: Retrieving Configuration-Related Code}} 
The main role of a \dev is to retrieve source code and conduct performance-aware code review, upon \perf's request, and respond with the result so that \perf has the necessary information to make the classification decision. 
Initially, \perf receives the potential performance-sensitive configuration to analyze. However, multiple methods across various classes may have some dependencies with the configuration parameter, making it difficult for \perf to assess the configuration's performance sensitivity accurately. Providing additional summaries of the configuration-related code, such as related source code and documentation, can help improve \perf's output~\cite{ye2020leveraging}. Hence, \perf relies on \dev, which utilizes two tools: (1) \textit{traditional program analysis} to extract source code that may be associated with the configuration through inter-procedural call graphs, and (2) \textit{document retrieval} to extract official documentation associated with the configuration. In addition to retrieving the code, \perf may also rely on \dev to provide feedback on the specific source code, since our intuition is that the developer should have a better understanding of the functionality of the code. Thus, \dev conducts performance-aware code reviews on the source code methods requested by \perf, as indicated in Figure~\ref{fig:template1}. 
Below, we further discuss how we extract code context, official documentation, and the prompt design for \dev's performance-aware code review. 

\noindent{\textit{\textbf{Extracting Configuration-Related Code.}}} We define configuration-related code as the caller source code that invokes a method that \textit{directly accesses the configuration}. For example, as indicated in Figure~\ref{fig:background}, the configuration under analysis is \texttt{key\_cache\_size\_in\_mb}, and its related source code is \texttt{initKeyCache}, which is the caller source code that accesses the configuration. To extract configuration-related source code, we first utilize static code analysis to extract the inter-procedural call graph~\cite{callGraph}. We first identify the method that directly accesses the configuration, then we traverse the graph to retrieve all the methods that have either a direct or indirect caller-callee relationship. % which are then parsed to extract the configuration-related code.

\noindent{\textit{\textbf{Extracting Configuration Documentation.}}} The description of a configuration on the document may provide additional information that can help classify configurations. For example, in the studied system Batik, the documentation for the configuration called \texttt{Width} provides additional information that it is the \textit{``Output Image Width"}, which may help the agents with the analysis. Therefore, we extract the configuration descriptions, if they are available, from the official project website. The description is passed to both \dev and \perf as part of the prompts when analyzing the configuration.

%We resort to manually extracting the official documentation since such documentation is not readily available or not even provided by some studied systems. provided by \texttt{.yaml} files only exists for the studied systems H2 and Cassandra. Moreover, our manual validation shows that not all studied systems have official documentation in their codebase. 
%Some systems may have poor documentation that reveals little about a configuration's details. For example, in the studied system Batik, the documentation for the configuration called \texttt{Width} is \textit{``Output Image Width,"} which provides very little information about its performance implications. In these cases, we find that including such documentation may degrade the quality of the LLM response~\cite{shi2023large}; hence, we augment the documentation with performance-related code.

\noindent{\textit{\textbf{\dev's Performance Aware Code Review.}}}
Figure~\ref{fig:template1} shows our prompt design for performance-aware code review that \dev carries out. Firstly, we give personification to the \dev, describing its role and goals, such as \textit{``You are a developer. Your job is to conduct performance-aware code review.''}. Consequently, we provide context about the (1) source code and (2) configuration description to the \dev. Finally, we ask \dev to output the following requirements: (i) summarize the functionality of the code, (ii) how many times such source code may be triggered (estimation based on the provided textual information), and (iii) whether the code may have an impact on memory or execution time. It is important to note that performance-aware code review does not determine whether a configuration is performance-sensitive; this task falls to \perf. However, \dev should be aware of common performance issues within the code they write, such as excessive memory usage and frequency of invocation, which may help \perf with the analysis. 

\begin{figure}
    \vspace{-3mm}
  \centering
  
  \begin{tcolorbox}[title={Prompt Template for Performance-Aware Code Review}]
    \textcolor{orange}{\textbf{Role:}} You are a developer. Your job is to conduct performance-aware code reviews on the given configuration-related code and official documentation for configuration to output the performance impact code that you wrote. \\
    \textcolor{deepgreen}{\textbf{Configuration-related code:}} 
    \begin{lstlisting}[language=Java]
private static AutoSavingCache initKeyCache()){
    ...
    long keyCacheInMemoryCapacity = DatabaseDescriptor.getKeyCacheSizeInMB() * 1024 * 1024;
    kc = CaffeineCache.create(keyCacheInMemoryCapacity);
    ...
}
\end{lstlisting}

    \textcolor{blue}{\textbf{Configuration description:}} Configuration Documentation after summarization. \\
        \underline{AutoSavingCache}: \textit{``Specify the way Cassandra allocates and manages memtable memory."} \\
    \textcolor{red}{\textbf{Requirement:}} You must output three things below: 
    
    1. Understand the functionality of the configuration in the code. 
    
    2. Investigate triggering frequency of configuration-related operations.
    
    3. Check the potential impact of configuration on the system.
  \end{tcolorbox}
    \vspace{-3mm}
  \caption{\dev's Performance-Aware Code Review.}
  \label{fig:template1}
    \vspace{-3mm}
\end{figure}

\subsubsection{\textbf{\underline{Performance Expert Agent}: Analyzing the Performance Sensitivity of Configuration}}

Given \dev's feedback on a specific configuration-related operation, \perf utilizes this feedback to classify performance-sensitive configurations. However, \perf may require further clarification on the retrieved code. For example, as indicated in Figure~\ref{fig:template2}, it may reference other methods (e.g., \texttt{create}) about which \perf may lack performance knowledge. Hence, \perf may request additional information about these operations. In particular, we use the prompt template in Figure~\ref{fig:template2}, which starts by personifying \perf with the introduction, \textit{``You are a performance expert... Check whether the provided configuration-related code is sufficient for performance analysis."} In the prompt, \perf receives the source code, as well as feedback from \dev as indicated in the template from Figure~\ref{fig:template1}. Based on this context, \tool instructs \perf to pinpoint unclear or ambiguous methods crucial for accurate performance analysis. Upon identifying the code that needs further analysis, \perf requests \dev to retrieve and analyze it. By retrieving and clarifying the code when needed, \perf can explore configuration-related code information, ensuring that all necessary code information is retrieved while minimizing the tokens and not exceeding the size limitation. % are retrieved for optimal performance classification.

\begin{figure}[h]
  \centering
  \vspace{-3mm}
  
  \begin{tcolorbox}[title={Prompt Template for Code Understanding}]
    \textcolor{orange}{\textbf{Role:}} You are a performance expert. Your job is to analyze the performance of the configuration. Check whether the provided configuration-related code is sufficient for performance analysis.  \\
    \textcolor{deepgreen}{\textbf{Configuration-related code:}} 
    \begin{lstlisting}[language=Java]
private static AutoSavingCache initKeyCache()){
    ...
    long keyCacheInMemoryCapacity = DatabaseDescriptor.getKeyCacheSizeInMB() * 1024 * 1024;
    kc = CaffeineCache.create(keyCacheInMemoryCapacity);
    ...
}
\end{lstlisting}

    \textcolor{blue}{\textbf{Code Context:}} Responses received from \dev. \\
    \textcolor{red}{\textbf{Requirement:}} If you need further code context to help understand the code, return the name of method name.
  \end{tcolorbox}
  \vspace{-3mm}
  
  \caption{\perf's Prompt for Code Understanding.}
  \label{fig:template2}
  \vspace{-3mm}
\end{figure}

\subsection{Multi-Agent Communications} 
Based on our definition of \dev and \perf, below we discuss how the agents collaborate together to classify performance-sensitive configurations.
\subsubsection{\textbf{\underline{Prompt Chaining} to Iteratively Build Code Understanding}}
One effective technique for enhancing the reliability and performance of LLMs is to use a prompting paradigm called prompt chaining. Prompt chaining refers to breaking a complex task into simpler subtasks, prompting the LLM with each subtask sequentially, and using its responses as inputs for subsequent prompts~\cite{PromptChaining}. In our performance chaining analysis, our goal is to retrieve all the necessary code for \perf to assess the configuration's sensitivity to performance. To achieve this, \perf iteratively instructs \dev to retrieve source code methods sequentially. The \dev fetches a single method based on \perf's requests (include the source code, \dev's description of the code, and \dev's performance-aware code review result) until a termination condition is met, indicating that \perf has gathered sufficient code and no longer requires assistance from \dev. \tool includes a memory mechanism that saves the \dev feedback at the end of each iteration of source code retrieval. This saved feedback can then be used as a code example in the next iteration of prompt chaining, allowing \perf to clarify unclear contexts and request additional source code methods if needed.

\subsubsection{\textbf{\underline{Retrieval Augmented Generation} for Performance Classifier}} 
\begin{figure}
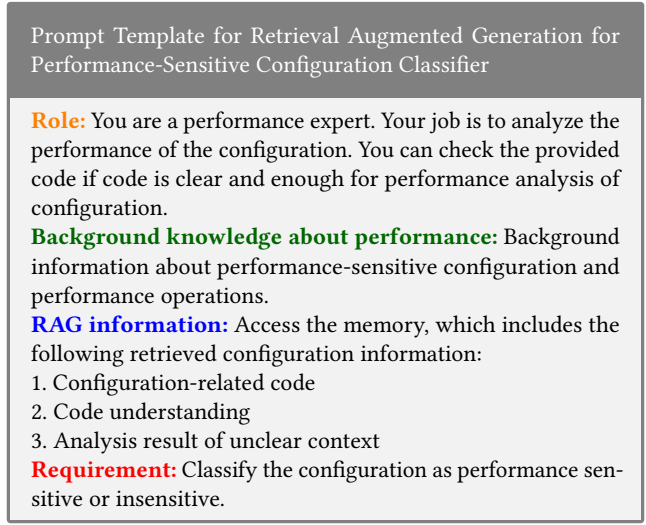

  \centering
  \vspace{-3mm}
  
  \begin{tcolorbox}[title={Prompt Template for Retrieval Augmented Generation for Performance-Sensitive Configuration Classifier}]
    \textcolor{orange}{\textbf{Role:}} You are a performance expert. Your job is to analyze the performance of the configuration. You can check the provided code if code is clear and enough for performance analysis of configuration. \\
    \textcolor{deepgreen}{\textbf{Background knowledge about performance:}} Background information about performance-sensitive configuration and performance operations.\\
    \textcolor{blue}{\textbf{RAG information:}} Access the memory, which includes the following retrieved configuration information: \\
    1. Configuration-related code \\
    2. Code understanding \\
    3. Analysis result of unclear context \\
    \textcolor{red}{\textbf{Requirement:}} Classify the configuration as performance sensitive or insensitive.
  \end{tcolorbox}
  \vspace{-3mm}
  \caption{Prompt Template 3: Retrieval Augmented Generation for Performance Classifier}
  \vspace{-3mm}
  \label{fig:template3}
\end{figure}

Based on the result of prompt chaining in prior steps, \perf sequentially builds a memory of the knowledge base, which allows \perf to classify performance-sensitive configurations more accurately. More precisely, we use the prompt template in Figure~\ref{fig:template3}. Like prior templates, our RAG starts by personifying \perf with the introduction, \textit{``You are a performance expert. Given feedback from \dev, your job is to perform performance analysis of configurations."} We then provide the retrieved context from \dev: (1) configuration-related code, (2) performance-aware code reviews, and (3) other code contexts to resolve clarity issues related to the configuration. Finally, we require \perf to classify whether or not the configuration is performance-sensitive.

\subsection{Implementation and Experiment Settings}

\phead{Environment.} We use GPT 3.5 (version gpt-3.5-turbo-0125) as our underlying LLM due to its popularity and wide usage. We leverage the OpenAI APIs and the LangGraph library~\cite{langlgraph} to implement the LLM agents for recursive code analysis and performance configuration classification. Temperature is a parameter in LLMs that ranges from 0 to 1. A low temperature makes the results more deterministic, and a higher value makes the results more diverse. To ensure the generated outputs are more stable across runs, We set the temperature to 0.3, which is a relatively low value but it still allows some diversity in the output. 
We also repeat our experiments five times and report the average. 
Note that although we use GPT 3.5 as the underlying LLM, our approach is general and can be replaced with other LLMs. 

\phead{Benchmark Datasets.} Table~\ref{tab:systems} presents the studied systems in our experiment. These seven systems are real-world open-source Java applications that cover various domains, ranging from databases to rendering engines. The systems have various configurations, some of which are related to performance. 
Previous work~\cite{10.1145/3611643.3616300} conducted manual performance testing and provided the ground truth about the performance-sensitive configurations for these seven systems. 
We leverage the ground truth provided by Chen et al.~\cite{10.1145/3611643.3616300} with some adjustments based on manually examining the source code and official documents. The replication package is available online~\cite{repo}.

\begin{table}
    \caption{An overview of the systems, versions, the number of configurations, and the number of performance-sensitive configurations that we studied.}    \label{tab:systems}
\centering
    \scalebox{0.9}{
    
    \begin{tabular}{l|l|l|r|r|}
         \toprule
         System & Domain & Version & Config. & Perf. \\ 
         &&&&Config.\\
         \midrule
         Cassandra & NoSQL Database & 4.0.5 & 133 & 76\\
         DConverter & image Density Converter & bdf1535  & 23 & 5\\
         Prevayler  & Database  & 2.6 & 12 & 8\\
         BATIK & SVG rasterizer & 1.14 & 21 & 8\\ 
         Catena & Password hashing & 1281e4b & 12 & 6 \\ 
         Sunflow & Rendering engine & 0.07.2 & 6 & 4\\    
         H2 & Database & 2.1.210   & 20 & 11\\
         \bottomrule
    \end{tabular}}
   % \begin{tablenotes}
   % \footnotesize
   % \item[] Opt: the number of options
   % \item[] Popt: the number of performance-sensitive option
   % \end{tablenotes}
\end{table}
%\phead{Evaluation Metric}
\section{Evaluation} \label{sec:evaluate}

In this section, we evaluate \tool by answering three research questions (RQs). 

%In this section, we conduct experiments
\subsection*{RQ1: How effective is \tool in identifying performance-sensitive configurations?}
%\phead{Motivation.} 

In this RQ, we evaluate the classification result of \tool in identifying the performance-sensitive configuration. %The evaluation can provide insights into the capabilities of \tool and its potential for studying configuration performance.
We compare \tool with two baselines: \DiagConfig and \ChatGPT. 
\DiagConfig~\cite{10.1145/3611643.3616300} utilized the taint static analysis on several systems to extract the performance-related operations related to configurations. Through manual performance tests by altering the configuration values and evaluating the variation of throughput, the performance-sensitive configurations would be identified and labeled. Utilizing the labeled configurations and taint static analysis of configuration, \DiagConfig is trained using a random forest model to classify performance-sensitive configurations.  
\ChatGPT directly calls ChatGPT APIs to classify if a configuration is performance-sensitive. We provide the system name, the configuration name, and the definition of a performance-sensitive configuration to ChatGPT (the same version as \tool) for classification. 
%\peter{then why do you give the name to chatgpt? we need to mention this in threat and approach about what we have done to reduce data leakage issue}
It is important to note that for \tool, the system name is not provided to reduce potential data leakage issues~\cite {10.1145/3639476.3639764}. %simulate a realistic software development setting where ChatGPT might not have updated knowledge of new or closed-source projects. 

The classification result of \tool is assessed using three accuracy metrics: accuracy, precision, and recall. 
Specifically, we focus on the precision and recall of classifying the performance-sensitive configurations. True Positives (TP) happen when a performance-sensitive configuration is correctly classified. True Negatives (TN) happens when a non-performance-sensitive configuration is correctly classified as not performance-sensitive. False Positives (FP) happen when \tool incorrectly classifies a non-performance-sensitive configuration as performance-sensitive. False negatives (FN) happen when \tool misclassifies a performance-sensitive configuration as non-performance-sensitive. %: Both \tool and the baseline classify the configuration as performance-sensitive.
Given the TP, FP, and FN, we calculate precision as $\frac{TP}{TP+FP}$ and recall as $\frac{TP}{TP+FN}$. Finally, accuracy measures the overall correctness of the classification of performance-sensitivity of configurations and is calculated as $\frac{TP+TN}{TP+TN+FP+FN}$. 

%The formulas for the calculation of accuracy, precision, and recall are as follows. Accuracy measures \tool’s ability to correctly classify all configuration candidates, precision focuses on accurately classifying configurations that are performance-sensitive, and recall evaluates \tool’s ability to identify all relevant performance-sensitive configurations.

\begin{comment}
\begin{itemize}
    \item[] 
    \begin{equation}
    \textbf{Precision} = \frac{TP}{TP+FP}
    \end{equation}

    \item[] 
    \begin{equation}
    \textbf{Recall} = \frac{TP}{TP+FN}
    \end{equation}

    \item True Positive (TP): Both \tool and the baseline classify the configuration as performance-sensitive.
    \item False Positive (FP): \tool incorrectly identifies a configuration as performance-sensitive.
    \item False Negative (FN): \tool misclassifies a performance-sensitive configuration as not performance-sensitive.
\end{itemize}
\end{comment}

Because of the generative nature of LLMs, the output may vary in each execution. Hence, we repeat each experiment five times and report the average precision, recall, and accuracy.

\begin{table*}
    \centering
    \caption{The accuracy, precision, and recall of \tool and the baselines in classifying performance-sensitive configurations. Note that \DiagConfig's results are unavailable for all the systems, except Cassandra, because \DiagConfig trained a classifier using other systems and applied it on Cassandra.}
    \scalebox{0.9}{
    \begin{tabular}{l|r|r|r|r|r|r|r|r|r|}
         \toprule
             & \multicolumn{3}{c|}{\textbf{\tool}} & \multicolumn{3}{c|}{\textbf{ChatGPT}} & \multicolumn{3}{c|}{\textbf{DiaConfig}} \\
              & Accuracy& Precision & Recall& Accuracy & precision& recall & Accuracy & precision & recall \\\midrule
         Cassandra & 64.01\%& 64.46\%& 82.32\%& 56.99\% & 57.08\% & 99.74\% &61.75\% &87.88\%& 38.26\%\\
         DConverter &66.09\%&39.06\% & 100.00\%& 26.96\% & 20.79\% & 84.00\% & --& -- & -- \\
         Prevayler & 75.00\%& 75.51\% & 95.50\% & 66.70\% & 66.70\%& 100.00\% & --&--& -- \\
         BATIK & 72.38\% & 63.41\% & 65.00\% & 34.29\% & 35.35\%  & 87.50\%&--& -- & --  \\ 
         Catena & 46.67\% & 48.15\% & 86.67\% & 50.00\% & 50.00\% & 83.00\%&--&--  & --  \\ 
         Sunflow & 53.30\% & 61.54\% & 80.00\% &66.70\% & 66.70\% & 100.00\%&--& -- & --  \\    
         H2 &76\% & 78.18\% & 78.18\% & 50.91\% & 50.46\%& 100.00\%& -- & -- & -- \\
         \midrule
         Average & 64.77\% & 61.47\% & 83.95\% & 50.36\% & 49.58\% & 93.46\% & -- & -- & --\\
         \bottomrule
    \end{tabular}}
    \label{tab: rq1}
\end{table*}

%\phead{Approach.} Previous work~\cite{10.1145/3611643.3616300} propose the framework, \texttt{DiagConfig} to classify the performance-sensitive configurations and provide a dataset of performance-sensitive options for the systems listed in Table~\ref{tab: study system}. We utilize this dataset and the same configurations as \texttt{DiagConfig} to evaluate the effectiveness of \tool. To assess the performance of \tool, we introduce the \texttt{ChatGPT baseline} for comparison. Specifically, we provide the system name, the configuration name, and the definition of a performance-sensitive configuration to ChatGPT with the same version to classify the performance sensitivity of configurations. It is important to note that for \tool, the system name is not provided to simulate a realistic software development setting where ChatGPT might not have updated knowledge of new or closed-source projects. 

%The performance of \tool is assessed using three accuracy metrics: accuracy, precision, and recall. Accuracy measures \tool’s ability to correctly classify all configuration candidates, precision focuses on accurately classifying configurations that are performance-sensitive, and recall evaluates \tool’s ability to identify all relevant performance-sensitive configurations. 

\phead{Results.} \textbf{\tool achieves a better accuracy (64.77\% on average) compared to \ChatGPT and \DiagConfig (50.36\% and 61.75\%, respectively).} Table~\ref{tab: rq1} shows the classification result of \tool and the baselines. 
We find that the \tool provides a better balance of precision and recall, achieving better accuracy than the two baselines. \ChatGPT achieves a higher recall (93.46\%) than both \tool (83.95\%) and \DiagConfig (38.26\%) but with a much lower precision (49.58\% v.s. 61.47\% and 87.88\%). We find that the reason for a high recall and low precision is that \ChatGPT misclassifies most configurations as performance-sensitive. In systems with less performance-sensitive configurations, \ChatGPT achieves much worse results. 
For example, in DConverter, since 80\% of the configurations are not performance-sensitive, \ChatGPT achieves only a 27\% accuracy rate. In contrast, the agents and prompting techniques implemented in \tool help improve the balance between precision and recall, resulting in much higher accuracy. 
We find that \DiagConfig achieves a relatively high precision of 87.88\% in Cassandra (it uses a classification model trained using data from all other systems, so the results are only available for Cassandra). However, \DiagConfig has a very low recall (38.26\% compared to \tool's 82.32\%) because \DiagConfig misses many configurations where there were issues with obtaining the taint analysis result. 

Note that, in theory, we can adjust \tool's precision/recall by asking LLMs to estimate the performance impact (e.g., severe, medium, or low) and only classify the ones with a severe impact as performance sensitive to improve precision. In our pilot study, we achieve a much higher precision of 72.41\% but a lower recall of 27.63\%. \tool only classifies the configurations with severe performance impact as the performance-sensitive configurations. However, in this work, we aim to achieve higher recall and maintain good precision because the goal of \tool is to provide an initial list of configurations for performance engineers efficiently for further investigation.

Compared to \DiagConfig, \tool requires less running time and no manual effort. \DiagConfig requires a taint analysis to identify all the code that is reachable from the configuration, which may require tens of hours of computation time for large systems like Cassandra. Moreover, \DiagConfig built a random forest classification model by manually collecting test results from other systems. This manual-intensive process may need to be repeated if we want to apply the model to systems in other domains or if they are developed by a different development practice. 
In contrast, \tool's running time is less than 50 minutes for Cassandra (the largest studied system with over 130 configuration parameters) and can be easily extended to any system because of its zero-shot and unsupervised nature.

\begin{tcolorbox}
\textbf{Answers to RQ1.} \tool provides a better balance of precision and recall, achieving the highest accuracy compared to the baselines. \tool is also lightweight and requires less than one hour to run for the largest studied system. %can accurately identify performance-sensitive configurations with smaller overhead, achieving a precision as high as 78\%
\end{tcolorbox}

\subsection*{RQ2: How do different components in \tool affect the classification result?}

\tool contains various components, including the retrieval augmented generation (e.g., retrieving related code to help make classification decisions) and chain-of-thought (e.g., asking agents to generate a code summary and combine the generated summary with subsequent tasks). In this RQ, we aim to study the impact of each component. We remove each component separately, re-execute \tool, and re-evaluate the
classification accuracy. In particular, we consider five combinations: 1) \texttt{code:} retrieve only the source code method that \textit{directly} uses the configuration value; 2) \texttt{code + analysis:} expands \texttt{code} by enabling the \dev to iteratively traverse the code to analyze the methods that the agent believes is relevant (i.e., through prompt chaining); 3) \texttt{dev:} the \dev generates a summary and description of the retrieved code; 4) \texttt{code + dev:} expands \texttt{code} by asking the \dev to provide a summary and description of the retrieved code for subsequent prompts (i.e., chain-of-thought); and 5) \texttt{code + dev + analysis:} the full version of \tool. %This component is about the developer's understanding of the retrieved code. \peter{what are they, list them}. Understanding the distinct role and impact of each component on the classification of performance sensitivity is essential. This detailed examination will determine the impact of each component and identify potential areas for enhancement in future LLM-based code comprehension research. 

\begin{table}
    \centering
    \caption{Classification results of \tool with different components. The best precision and recall values in each system are marked in bold. The numbers in the parentheses show the percentage difference compared to the full version of \tool (\texttt{code+dev+analysis}). }
    \label{tab:rq2}
    \scalebox{0.78}{
    \begin{tabular}{llll}
         \toprule
         System & Approach & Precision & Recall\\ \midrule
         \multirow{5}{*}{\rotatebox[origin=c]{90}{Cassandra}}
         &\textbf{\tool}$_{\text{code+dev+analysis}}$ & 64.46\% & \textbf{82.32\%} \\
         &\textbf{\tool}$_{\text{code}}$ & \textbf{67.55\% (+4.79\%)} & 73.42\% (-10.81\%) \\
         &\textbf{\tool}$_{\text{code+analysis}}$&61.81\% (-4.11\%)& 77.11\% (-6.32\%)\\
         &\textbf{\tool}$_{\text{dev}}$&62.32\% (-3.31\%) & 74.47\% (-9.53\%)\%\\
         &\textbf{\tool}$_{\text{code+dev}}$& 64.12\% (-0.53\%) \%&73.09\% (-11.23\%)\\ \midrule
         \multirow{5}{*}{\rotatebox[origin=c]{90}{Dconverter}}
         &\textbf{\tool}$_{\text{code+dev+analysis}}$& \textbf{39.06\%} & \textbf{100.00\% }\\    
         &\textbf{\tool}$_{\text{code}}$ & 30.49\% (-21.94\%)  & 100.00\% (-0.00\%) \\
         &\textbf{\tool}$_{\text{code+analysis}}$& 29.76\% (-23.81\%) & 100.00\% (-0.00\%)\\
         &\textbf{\tool}$_{\text{dev}}$& 32.20\% (-17.56\%)& 100.00\% (-0.00\%)\\
         &\textbf{\tool}$_{\text{code+dev}}$& 36.76\% (-5.89\%)  & 100.00\% (-0.00\%)\\ \midrule
         \multirow{5}{*}{\rotatebox[origin=c]{90}{Prevayler}}
         &\textbf{\tool}$_{\text{code+dev+analysis}}$& 75.51\% & \textbf{92.50\%}\\
         &\textbf{\tool}$_{\text{code}}$ & \textbf{80.56\% (+6.70\%) }& 72.50\% (-21.62\%)\\
         &\textbf{\tool}$_{\text{code+analysis}}$& 74.47\% (-1.38\%)  & 87.50\% (-5.41\%) \\
         &\textbf{\tool}$_{\text{dev}}$& 70.83\% (-6.20\%)& 85.00\% (-8.11\%) \\
         &\textbf{\tool}$_{\text{code+dev}}$& 70.45\% (-6.70\%) & 79.49\% (-14.06\%)\\ \midrule
         \multirow{5}{*}{\rotatebox[origin=c]{90}{BATIK}} 
         &\textbf{\tool}$_{\text{code+dev+analysis}}$& \textbf{63.41}\% & \textbf{65.00\%}\\
         & \textbf{\tool}$_{\text{code}}$& 42.86\% (-32.40\%) & 45.00\% (-30.76\%)\\
         &\textbf{\tool}$_{\text{code+analysis}}$&47.92\% (-24.42\%) & 57.50\% (-11.53\%) \\
         &\textbf{\tool}$_{\text{dev}}$& 59.09\% (-6.86\%) & 32.50\% (-50.00\%)\\
         &\textbf{\tool}$_{\text{code+dev}}$& 51.28\ (-19.12\%) & 50.00\% (-23.07\%)\\ \midrule
         \multirow{5}{*}{\rotatebox[origin=c]{90}{Catena}}
         &\textbf{\tool}$_{\text{code+dev+analysis}}$&48.15\% & \textbf{86.67}\%\\
         &\textbf{\tool}$_{\text{code}}$ & \textbf{51.43\% (+6.81\%)} & 60.0\% (-30.77\%)\\
         &\textbf{\tool}$_{\text{code+analysis}}$&48.89\% (+1.53\%)  & 73.33\% (-15.39\%) \\
         &\textbf{\tool}$_{\text{dev}}$& 48.15\% (-0.00\%) & 86.67\% (-0.00\%) \\
         &\textbf{\tool}$_{\text{code+dev}}$& 50.00\% (+3.84\%) & 76.67\% (-11.53\%) \\ \midrule
         \multirow{5}{*}{\rotatebox[origin=c]{90}{Sunflow}}
         &\textbf{\tool}$_{\text{code+dev+analysis}}$& 61.54\% & 80.00 \%\\ 
         &\textbf{\tool}$_{\text{code}}$ & \textbf{83.00\% (+34.87\%)} & 50.00\% (-37.50\%)\\
         &\textbf{\tool}$_{\text{code+analysis}}$& 65.22\% (+6.00\%)& 75.00\% (-6.25\%)\\
         &\textbf{\tool}$_{\text{dev}}$& 68.00\% (+10.49\%)& \textbf{100\% (+25.00\%)} \\
         &\textbf{\tool}$_{\text{code+dev}}$& 65.38\% (+6.23\%) & 89.47\% (+11.83\%)\\\midrule
         \multirow{5}{*}{\rotatebox[origin=c]{90}{H2}}
         &\textbf{\tool}$_{\text{code+dev+analysis}}$& 78.18\% & \textbf{78.18\%} \\ 
         &\textbf{\tool}$_{\text{code}}$ & 58.70\% (-24.91\%)& 50.00\% (-36.04\%)\\
         &\textbf{\tool}$_{\text{code+analysis}}$& 66.04\% (-15.52\%) & 63.64\% (-18.59\%)\\
         &\textbf{\tool}$_{\text{dev}}$& 66.07\% (-15.48\%)& 67.27\% (-13.95\%) \\
         &\textbf{\tool}$_{\text{code+dev}}$& \textbf{80.00\% (+2.33\%)} & 72.73\% (-6.97\%)\\\midrule
    \end{tabular}}

\end{table}

\phead{Results.} \textbf{Integrating \texttt{code} results in the highest precision in 4/7 studied systems, with the sacrifice of recall. Further adding \texttt{dev} improves recall significantly across most systems, achieving a more balanced trade-off between precision and recall while maintaining reasonably high precision.} Table~\ref{tab:rq2} shows the precision and recall of the combinations of the components in \tool. We find that a simple RAG approach by retrieving only the method that directly uses the configuration parameter (\tool$_{\text{code}}$) can achieve good precision across all studied systems but a much lower recall compared to the full version of \tool (12.00\% to 36.04\% lower). For instance, in the case of Cassandra, \tool$_{\text{code}}$ achieves a precision of 67.55\%, which is 4.79\% higher than the full version, but the recall drops significantly by 10.81\% to 73.42\%.

While \texttt{code} alone can achieve high precision, it often sacrifices recall significantly. On the other hand, \texttt{code+dev} tends to provide a better balance between precision and recall, with generally higher recall rates and more stable precision. This suggests that incorporating LLM-generated summaries and descriptions helps to enhance the overall performance of the tool by maintaining a more comprehensive approach. 
In comparison, integrating the LLM-generated summary/description of a given piece of code in the prompt (\texttt{code} v.s. \texttt{code+dev}) tends to provide a better balance between precision and recall, with generally higher recall rates and more stable precision. This suggests that incorporating LLM-generated summaries/descriptions helps to enhance the overall performance of \tool by maintaining a more comprehensive approach. %we see a general increase in recall but a mixed effect in precision (decrease in 4/7 systems). The results show that LLMs are more likely to be able to make a decision 

\noindent {\bf Adding \texttt{analysis} to \texttt{code} further improves recall significantly across most systems while maintaining similar precision.} For example, in Prevayler, the precision of \texttt{code+analysis} (74.47\%) is slightly lower than \texttt{code} (80.56\%), but the recall increases from 72.56\% to 87.50\%. Similarly, in BATIK, while \texttt{code+analysis} achieves a precision of 47.92\% compared to 42.86\% for \texttt{code}, the recall improves significantly from 45.00\% to 57.50\%. This finding suggests that the additional context and understanding provided by the analysis help \tool identify more relevant methods, thereby improving recall and maintaining precision. %These examples highlight that while precision might slightly decrease, the overall recall benefits substantially from the inclusion of analysis.

\noindent \textbf{Integrating \texttt{code+dev+analysis} offers a holistic approach that leverages the strengths of individual components—code, developer insights, and analytical context—to achieve the best balanced performance.} For instance, in the case of Cassandra, \texttt{code+dev+analysis} achieves a precision of 64.46\% and a recall of 82.32\%. The full version of \tool surpasses the recall of both \texttt{code} (73.42\%) and \texttt{code+dev} (73.09\%), while maintaining a competitive precision. 
Our findings show that each component has its own benefits to the result, and the integrated components enhance the overall effectiveness and reliability of \tool, providing a robust solution for identifying relevant methods and classifying performance-sensitive configurations within codebases.

\begin{tcolorbox}
\textbf{Answers to RQ2.} Adding \texttt{dev} improves recall with a balanced trade-off in precision, and incorporating \texttt{analysis} further enhances recall while maintaining competitive precision. The combined \texttt{code+dev+analysis} approach effectively leverages each component's strengths for comprehensive method identification.

%The integration of source code information can enhance precision, while developer analysis may boost recall. Effective collaboration between these elements typically yields superior results.
\end{tcolorbox}

\subsection*{RQ3: What are the reasons for \tool's misclassification?} \label{rq3}
Despite \tool achieving the highest accuracy and maintaining a balance between precision and recall, there remain instances of misclassification. Understanding the underlying causes of these misclassifications is crucial for understanding the limitations of \tool and providing insights for future performance analysis utilizing LLMs. Hence, in this RQ, we conduct a detailed manual study on the reasons for misclassification. 

We collected and examined 362 configurations incorrectly classified by \tool. To systematically analyze the reasons for misclassifications by \tool, we began by selecting a 20\% random sample from our dataset of 362 misclassified configurations. This subset was thoroughly reviewed to identify and categorize the various reasons for misclassification. In particular, we studied the communication history among the agents, the source code, and all related documents that we could find. With these categories established, we then manually examined the remaining 80\% of the dataset, applying the derived categories to each configuration to understand the distribution of the different reasons for misclassification. We did not find any new categories during the process.

%\wang{not sure if we have modifications on the classification but I have gone through and labeled misclassification, and it is easy for me to do modifications on classifications.}

\begin{table*}
    \centering
    \caption{Reasons and prevalence for the misclassification of performance sensitivity by \tool. }
    \scalebox{0.9}{
    %\resizebox{\textwidth}{!}{
    \begin{tabular}{llll}
         \toprule 
          Reason & Description & Percentage  \\ \midrule

          \parbox[c]{0.25\linewidth}{No clear evidence of performance sensitivity } & \parbox[c]{0.6\linewidth}{Through an examination of the related code and a careful review of available information (e.g., code and documentation), there is no clear evidence to indicate the performance sensitivity of the configuration.
          % he available performance-related information is limited, making it challenging to classify performance sensitivity.
         } 
         & 54.1\% \\ 
         \midrule
         Misunderstood requirements & \parbox[c]{0.6\linewidth}{LLMs misunderstand the requirements for classification of performance-sensitive.} & 26.8\% \\ \midrule  

         \parbox[c]{0.25\linewidth}{Incorrect interpretation on the impact of performance-related operation} &\parbox[c]{0.6\linewidth}{LLMs incorrectly interpreted the impact of performance-related operations. This misinterpretation led to the misclassification of the performance sensitivity of the configurations.}  &  10.0\%\\\midrule
         
         \parbox[c]{0.25\linewidth}{Influenced by performance-related keywords}&  \parbox[c]{0.6\linewidth}{When LLMs classifies performance-sensitive configurations, keywords like “memory” and “scalability” can lead to misclassifications. These keywords are inherently associated with performance-related aspects, which may cause performance-insensitive configurations to be incorrectly identified as performance-sensitive.} & 8.0\%\\ \midrule
         
         % Code impacts judgment & \parbox[c]{0.5\linewidth}{has some performance code but it is not sensitive. Or don't have clear time-expensive code but sensitive }& 28\%\\ \midrule 
         
        Hallucination & \parbox[c]{0.6\linewidth}{LLMs generate information that is not based on actual facts or truths. }& 1.1\% \\ 

         \bottomrule
    \end{tabular}}
    \label{tab: misclassification reason}
\end{table*}

\phead{Results} Table~\ref{tab: misclassification reason} shows the reasons for the misclassification of performance-sensitivity by \tool and their percentage. In total, we uncovered five reasons that cause the misclassification.

\phead{Most misclassifications (54.1\%) occur in configurations without clear evidence to support the performance sensitivity.} For the configuration where the classification results by \tool differ from the ground truth~\cite{10.1145/3611643.3616300}, although we conducted a thorough examination of the related code and a careful review of available information (e.g., documentation and source code), there is no substantive evidence supporting the performance sensitivity of this configuration. For example, in Cassandra, the configuration \texttt{column\_index\_cache\_size\_in\_kb} is not performance-sensitive~\cite{10.1145/3611643.3616300}. However, the LLM agents responded that this configuration can impact the amount of memory used for holding index entries in memory, which can cause performance variations. Setting a higher value may improve performance by reducing disk reads for index entries while setting a lower value may result in more disk reads and potentially slower performance. Based on reading the source code, we believe the explanation of \tool's decision is valid, but there is no available evidence to support the performance sensitivity of the configuration. 
This misinterpretation can lead to inaccurate classification of performance sensitivity, highlighting the need for providing better requirements to LLMs for code understanding. % assessments and misguided optimization efforts, highlighting the importance of precise and context-aware analysis in performance sensitivity classification.

\phead{\tool may misunderstand requirements on performance sensitivities and classify other aspects as performance sensitive (26.8\%).} \tool may do some interpolation and infer that some non-performance-sensitive operations as performance sensitive. For example, the configuration \texttt{\_prevalenceDirectory} in Prevayler specifies the directory used for reading and writing files. The configuration is related to file storage and is not performance-sensitive. However, \tool incorrectly assumes that the location of these read-write operations impacts system performance, whereas the configuration primarily pertains to system storage rather than performance.

%wrongly assert that the amount of logging generated during Garbage Collection events can impact system performance. Setting a lower threshold may lead to more frequent logging, resulting in increased I/O operations and log processing. However, adjusting the value of this configuration doesn’t change how garbage collection works or the efficiency of the database operations, and it only affects what information is logged. Therefore, it is not directly sensitive to performance, but rather a tool for identifying performance issues. 

%For instance, the configuration \texttt{ \_prevalenceDirectory} in the Prevayler project is not performance-sensitive based on the ground truth, and specifies the directory used for reading and writing files, which relates to system storage. However, LLMs incorrectly assert that the location of these read-write operations impacts system performance, but the location of file reading is related to the system storage rather than the system performance. 

\phead{10.0\% of the misclassifications are due to incorrect interpretations of performance impact.}
% \wang{ I feel the case in this category can be merged to no clear evidence. Because we don't know if the interpretation of the impact of performance-related operations is correct or not. Like ChatGPT said the creation of a new GenericSnapshotManager object in the Prevayler project is time-consuming but the baseline says it is not performance-sensitive. we don't know which one is correct. } 
% For example, ground truth~\cite{10.1145/3611643.3616300} indicates that the configuration \texttt{\_nullSnapshotManager} in Prevalyer is not performance-sensitive. However, \tool argues that this configuration determines whether a new \texttt{GenericSnapshotManager} object is created or if the existing one is returned, which has an impact on the system performance. 
In some instances, \tool inaccurately assesses the performance impact of specific configurations that do not inherently influence system efficiency.
For example, the configuration \texttt{BACKGROUND\_COLOR} in Batik, which sets the background color, is performance-sensitive. The reason is different color settings can have an impact on the performance of graph rendering. However, \tool incorrectly classifies the code related to the configuration of the color as performance-insensitive. %without any performance impact. However, different color settings in batik can have an impact on the performance of graph rendering \peter{citation}. 

\phead{8.0\% misclassification is related to having performance-related keywords for a performance non-sensitive configuration.} LLMs are trained using natural language texts so they are sensitive to keywords in the prompts. If there are performance-related keywords in the prompt, \tool is more likely to classify a configuration as performance-sensitive. 
%For example, the configuration \texttt{enableAntiAliasing} in Dconverter is not performance-sensitive based on the groundtruth\peter{cite}. Enabling or disabling the configuration does not affect the system execution time. However, because the keyword ``antialiasing'' is considered performance-related in many graphics-related projects, the \tool believes enabling antialiasing for image processing operations can have a notable impact on system performance and misclassify the configuration as performance-sensitive. 
For example, the configuration \texttt{gc\_log\_threshold\_in\_ms'} is not performance-sensitive in the Cassandra project based on the ground truth~\cite{10.1145/3611643.3616300}. Enabling or disabling the configuration does not affect the system execution time.  However, the keyword ``gc'' (garbage collection) is often considered to be performance-related in many situations, \tool incorrectly classified the configuration as performance-sensitive. However, the configuration is related to logging during gc, and setting a lower/higher threshold does not have a noticeable impact on the performance. 

% For example, the configuration \texttt{cross\_node\_timeout} is not performance-sensitive in the Cassandra project based on the benchmark dataset, LLMs think this configuration impacts the behavior of the \texttt{getExpiresAtNanos} method, which is crucial for determining operation timeout information exchange between nodes. Therefore, enabling or disabling this configuration affects the request timeouts between nodes and has a performance impact on the system running.

\phead{Hallucination is rare but can still cause misclassification (1.1\%).} Hallucination in LLMs can lead to incorrect or misleading results~\cite {liu2024exploring, li2024dawn}. For example, the configuration \texttt{hinted\_handoff\_enabled} in Cassandra is considered performance-sensitive~\cite{10.1145/3611643.3616300}. This configuration is to allow Cassandra to ``\textit{continue performing the same number of writes even when the cluster is operating at reduced capacity}''. However, due to hallucination, \tool erroneously states that this configuration is related to the name of applications, causing misclassification of the configuration.

\begin{tcolorbox}
\textbf{Answers to RQ3.} \tool's misclassifications of performance-sensitive configurations are primarily due to a lack of clear evidence supporting performance sensitivity (54.1\%) and misunderstanding of requirements leading to incorrect classifications (26.8\%). %, incorrect interpretation of performance impacts (13.3%), influence of performance-related keywords causing misclassification (11.4%), and rare occurrences of hallucination (1.1%). 
Addressing these issues may require better requirements specification and enhanced understanding by LLMs to improve classification accuracy.

% No clear evidence to support the performance sensitivity is the main reason for the misclassification of performance-sensitivity of configurations (45.3\%). 
\end{tcolorbox}

\section{Discussion} \label{sec:disscussion}

%\peter{in no particular order}
%\peter{1) need better requirement, may be part of the prompt}
%\peter{2) performance sensitivity is hard, \tool can help narrow down the configs that needs investigation quickly}
%\peter{3) describe the benefits of our prompting technique (prompt chaining and providing additional summary)}

\phead{Better requirements for analyzing the performance sensitivity of configuration are needed.} During the reason analysis of the misclassification of configurations, we find that misunderstanding the requirement on performance sensitivities affects the precision in identifying the performance-sensitive configurations by \tool. The specificity and clarity of prompts used to interact with these agents can influence their ability to accurately identify performance-sensitive configurations. Incorporating more explicit performance-related criteria into the prompts can help reduce misclassifications by aligning the LLM’s analysis more closely with the actual performance impacts of the configurations. This adjustment could guide the LLM agents to distinguish between configurations that truly affect performance and those that do not, despite potentially misleading indicators such as performance-related keywords. 

\phead{\tool efficiently narrows down the scope of investigation performance sensitivity of configurations.} One of the strengths of \tool is its ability to efficiently narrow down the list of configurations that need deeper investigation. Performance sensitivity in software configurations is a complex domain where manual identification processes are time-consuming and prone to errors. By automating the identification process, \tool allows performance engineers and developers to focus their efforts and expertise on a refined subset of configurations, enhancing productivity and optimizing resource allocation. 
%\peter{discuss the trade-off in precision and recall here}

\phead{Prompt chaining and RAG technique enhance \tool understanding and analytical capabilities of the LLM on performance analysis.} tool leverages advanced prompting techniques, such as prompt chaining and retrieval-augmented generation, to improve the interaction dynamics between the developer and performance expert agents. Prompt chaining breaks down complex tasks into simpler, sequential queries that build upon each other, which helps in constructing a comprehensive performance analysis for each performance-sensitive decision. RAG integrates the configuration-related information from external sources and reduces the context size, ensuring that the analysis from LLMs is both relevant and deeply informed for the performance assessment of configurations. The integration of various prompting techniques enhances \tool's ability to accurately identify performance-sensitive configurations with balanced precision and recall,  minimizing the risk of overlooking critical performance nuances in software behavior.

\section{Threats to Validity} \label{sec:threats}
\subsection{Internal Validity } Due to the generative nature of LLM, the responses may change across runs. To mitigate the threat, we
try to execute the LLMs five times and report the average for our evaluation. We set the temperature value to 0.3, which makes the result more consistent but still allows some diversity. We find that the results are similar across runs, which means the outputs are stable. However, future studies are needed to understand the impact of temperature on the results. Since LLMs are trained using open-source systems, there is the possibility of data leakage problems. To minimize the impact, we excluded system-specific information (e.g., system and package names) when classifying configuration performance sensitivity to mitigate data\cite{10.1145/3639476.3639764}. 

\subsection{External Validity }We conducted the study on open-source Java systems. Although we tried to choose matured and popular systems that are also used in prior studies, the results may not apply to systems implemented in other programming languages. Future research is needed to examine the results of other types of systems. %Therefore, 

\subsection{Construct Validity }Classifying the performance sensitivity of a configuration parameter is a challenging task due to varying workloads~\cite{vitui2021mlasp}. Hence, in this paper, we rely on the prior benchmark~\cite{10.1145/3611643.3616300} and validate the result by an in-depth analysis and all the documents that we could find. To encourage replication and validation of our study, we made the dataset publicly online~\cite{repo}. 
\section{Conclusion} \label{sec:conclusion}
Configuration parameters are crucial for customizing software behavior, and some configurations can have a performance impact on systems. However, misconfigurations are common and can lead to significant performance degradation, making it essential to identify performance-sensitive configurations. In this paper, we introduced \tool, a novel framework leveraging LLM-based multi-agent systems to identify performance-sensitive configurations in software systems. By combining static code analysis and retrieval-augmented prompting techniques, \tool can identify performance-sensitive configurations with minimal manual work. Our evaluation of seven open-source systems demonstrated that \tool achieves a higher accuracy of 64.77\% compared to existing the state-of-the-art method (61.75\%). Furthermore, our evaluation of studying the effect of different prompting components revealed that the implementation of prompt chaining in \tool substantially enhances recall, with improvements ranging from 10\% to 30\%. To understand the limitations of \tool, we conduct a manual analysis of 362 misclassification configurations to analyze and summarize the reasons for the misclassification of performance sensitivity by \tool. LLM’s misunderstanding of requirements (26.8\%) is the key reason for misclassification. Additionally, we provide a detailed discussion to offer insights for future research to enhance the robustness and accuracy of LLM-based configuration performance analysis.

\balance
\bibliographystyle{ACM-Reference-Format}
\bibliography{ref}
\end{document}